\pdfoutput=1
\documentclass[pdftex]{article}
\usepackage{amssymb}
\usepackage{icrctc07}

\title{Lateral distribution and the energy determination of showers along the ankle}
\shorttitle{Lateral distribution and energy}

\authors{G. Ros$^{1,2}$, G. A. Medina-Tanco$^{3}$, C. De
Donato$^{4}$, L. del Peral$^{1}$, D. Rodr\'iguez-Fr\'ias$^{1}$, J.C.
D'Olivo$^{3}$, J.F. Vald\'es-Galicia$^{5}$, F. Arqueros$^{2}$.}
\shortauthors{G. Ros and et al}

\afiliations{$^1$ Space Plasmas and Astroparticle Group, Universidad
de Alcal\'a, Pza. San Diego, s/n, 28801 Alcal\'a de Henares
(Madrid), Spain. \\ $^2$ Dpto. F\'isica At\'omica, Molecular y
Nuclear, Facultad de F\'isica, Universidad Complutense de Madrid,
Ciudad Universitaria 28040, Madrid (Spain). \\ $^3$ Dpto. Altas
Energ\'ias, Inst. de Ciencias Nucleares, Universidad Nacional
Aut\'onoma de M\'exico, M\'exico D.F., C.P. 04510. (M\'exico). \\
$^4$ Dipartimento di Fisica dell'Universit\'a degli Studi di Milano
and Sezione INFN, via Celoria 16, I-20133.
\\$^5$ Inst. de Geof\'isica. Universidad Nacional Aut\'onoma de M\'exico, M\'exico D.F.,
C.P. 04510.}

\email{german.ros@uah.es}

\abstract{The normalization constant of the lateral distribution
function (LDF) of an extensive air shower is a monotonous (almost
linear) increasing function of the energy of the primary. Therefore,
the interpolated signal at some fixed distance from the core can be
calibrated to estimate the energy of the shower. There is, somehow
surprisingly, a reconstructed optimal distance, $r_{opt}$, at which
the effects on the inferred signal, $S(r_{opt})$, of the
uncertainties on true core location, LDF functional form and
shower-to-shower fluctuations are minimized. We calculate the value
of $r_{opt}$ as a function of surface detector separation, energy
and zenith angle and we demonstrate the advantage of using the
$r_{opt}$ value of each individual shower instead of a same fixed
distance for every shower, specially in dealing with events with
saturated stations. The effects on the determined spectrum are also
shown.}

\begin{document}
\maketitle

\section{Introduction}

In order to determine the energy of cosmic rays with surface
detectors arrays, first the lateral distribution function (LDF) of
the shower particles, i.e. the particles density or signal versus
distance to shower core location, is fitted assuming a known
functional form. Following Hillas \cite{Hillas} proposal, the signal
at some fixed distance of the shower core $S(r_{opt})$ for all the
showers, independent of their energy or direction, is used to relate
it with primary energy, usually using monte carlo simulations. The
optimum distance $r_{opt}$ is mainly related to the geometry of the
array.

We show that this method may not reconstruct properly the shape of
the spectrum. We use an AGASA-like experiment \cite{AGASASpectrum}
as case study and inject both a single power law and a simplified,
yet realistic, structured spectrum above $10^{17.7}$ eV. A special
analysis has been done for saturated events.

\section{Method}

A previous version of our algorithm was presented in
\cite{GustavoAlgorithm}. We use a simplified numerical approach to
the simulation of extensive air shower detection in a surface array.
Our ideal detector is an infinite array of equally spaced stations
distributed in elementary triangular cells of 1 km side.

The input spectrum is a perfect power law spectrum with index -3.0
from $10^{18.0}$ to $10^{20.1}$ and with an isotropic zenith
distribution from 0 to 45 degrees. The number of events is 230.000,
approximately the same statistic as the spectrum reported by AGASA.
In a second case, we used a structured spectrum (with second knee,
ankle and GZK-cut-off, and exposure-limited at low energy) from
$10^{17.7}$ to $10^{21.0}$ eV as input.

Firstly, we select random core position inside an elementary cell,
and the signal of each station is estimated using the LDF reported
by Auger \cite{Auger}:

\vspace{-0.3cm}

\begin{eqnarray}\label{equ1}
S(r_{km},E_{EeV},\theta)=\frac{7.53\;E^{0.95}\;2^{\beta(\theta)}}{\sqrt{1+11.8[sec(\theta)-1]^2}} \nonumber \\
 \times r^{-\beta(\theta)}\times(1+r)^{-\beta(\theta)}
\end{eqnarray}

\noindent where $\beta(\theta)=3.1-0.7sec(\theta)$. The signal
expected at each station is then fluctuated by Poissonian noise and
recorded if it is above a threshold of 3.0 VEM (Vertical Equivalent
Muons, the signal deposited by one vertical muon in an Auger water
Cerenkov tank). Stations with a signal $S_{i} > S(0.2$ km $,1$ EeV
$,0^o)$ are considered saturated and are excluded.

The \textit{real} core position is now moved using a gaussian
distribution centered at this point. The sigma of this distribution
is set taking into account the uncertainty in core determination,
which depends on the array geometry and primary energy
\cite{M.C.Medina}. The new core position simulates the
\textit{reconstructed} core. In order to mimic the reconstruction
procedure we fit the signals of the triggered stations with the LDF
of the AGASA form:

\vspace{-0.3cm}

\begin{equation}\label{equ2}
\log S(r_{m})=a_{1}-a_{2}\log(r/r_{M})-0.6\log(1+r^{2})
\end{equation}

\noindent with $r_{M}=91.6$ m, the moliere radius at AGASA altitude.
Finally, the signal at 600 m ($S_\theta(600)$), is used to estimate
the primary energy.

For the same shower, another procedure is done to obtained its real
optimum distance $r_{opt}$. The \textit{reconstructed} core is
shifted 50 times using again a gaussian distribution centered at
this point and with the same sigma as before. For each new core
position, an LDF is fitted and the point $r_{opt}$ is defined as the
location of the minimum dispersion. With the signal at this point
$S(r_{opt})$, the energy of the primary is again estimated.

\subsection{Conversion between Auger and AGASA LDFs}

A conversion of units between the LDF from Auger (that we have used
to assign the signal in each station) and the LDF from AGASA (that
we will use for energy determination) is needed. The AGASA LDF is:

\vspace{-0.5cm}

\begin{eqnarray}\label{equ3}
\rho(r_m) = K \left(\frac{r}{r_{M}}\right)^{-1.2}  \nonumber \\
\times \left(1+\frac{r}{r_{M}}\right)^{-(\eta(\theta)-1.2)} \times
\left(1+r^{2}\right)^{-0.6}
\end{eqnarray}

where $\eta=3.84-2.15(sec(\theta)-1)$ and K is the shower size. The
conversion factor (AGASA-LDF/Auger-LDF), depends on energy, zenith
and core distance. A study of the energy and zenith dependence over
the hole spectral range shows that it is negligible. Nevertheless,
the dependence on core distance is sizable and a fit of the form
$1/(a+bx^c)$ is used.

\subsection{$r_{opt}$ dependence with array spacing}

The dependence of $r_{opt}$ with the energy, zenith and detectors
separation has been studied and it is presented in
\cite{GRosbienal}. Here we show the dependence with array spacing
for several primary energies and $\theta = 30$ deg (see Fig. 1). The
results are in agreement with the values used by AGASA (detectors
separated 1 km and $r_{opt} = 600$ m) and Auger (separation of 1.5
km and $r_{opt} = 1000$ m). Note, however, the dependence of
$r_{opt}$ with energy and its considerable dispersion.

\begin{figure}
  \begin{center}
    \includegraphics[width=0.48\textwidth]{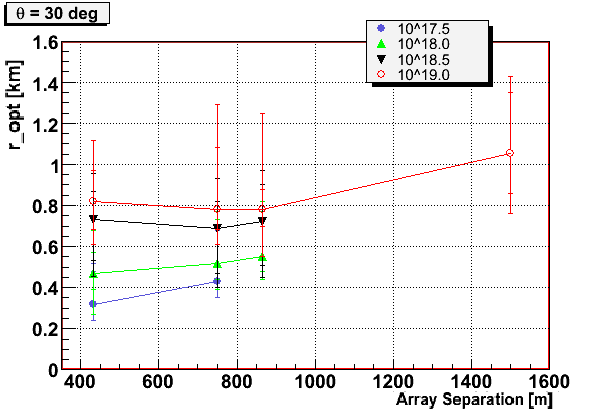}
  \end{center}
  \vspace{-0.5cm}
  \caption{$r_{opt}$ for different array spacings and primary
  energies at $\theta = 30$ deg. The error bars represent the CL at 68 and 95\%.}
\end{figure}

\subsection{Energy determination}

Since traversed atmosphere is a function of zenith angle, AGASA
experiment transformed the observed shower density $S_{\theta}(600)$
at zenith angle $\theta$ into $S_{0}(600)$, the corresponding value
of a vertical shower. The attenuation curve is:

\vspace{-0.5cm}

\begin{eqnarray}\label{equ4}
S_{\theta}(600)=S_{0}(600)\,f_{s}(\theta)= S_{0}(600) \nonumber \\
\times
\,exp\left[-\frac{X_0}{\Lambda_1}(sec\theta-1)-\frac{X_0}{\Lambda_2}(sec\,\theta-1)^2\right]
\end{eqnarray}

where $X_0=920\; g/cm^2$, $\Lambda_1=500\; g/cm^2$ and
$\Lambda_2=594\; g/cm^2$ for showers with $\theta<45$ deg. The
uncertainty in $S_{0}(600)$ due to this transformation is estimated
to be $\pm5\%$. The conversion formula to relate $S_{0}(600)$ with
energy reported by AGASA is:

\vspace{-0.3cm}

\begin{equation}\label{equ5}
E=2.21\cdot10^{17}S_0(600)^{1.03} eV
\end{equation}

\noindent where different hadronic interaction models and simulation
codes were considered. Using eq. \ref{equ5} we calculated the shower
energy based on the observed signal at 600 m.

In order to find the energy using the signal at $r_{opt}$, we use
the following parametrization of the shower size (obtained from eq.
\ref{equ3} and eq. \ref{equ5}):

\vspace{-0.5cm}

\begin{eqnarray}\label{equ6}
K(\theta,E_{EeV}) = 49.676\times f_s(\theta) \nonumber \\
 \times \left(1+\frac{600}{r_M}\right)^{\eta(\theta)-1.2}\times
E^{1/1.03}
\end{eqnarray}

Once the $S(r_{opt})$ has been determined as explained before, the
shower size is obtained from eq. \ref{equ3} and the energy from eq.
\ref{equ6}.

\section{Results and discussion}

\begin{figure}
  \begin{center}
    \includegraphics[width=0.48\textwidth]{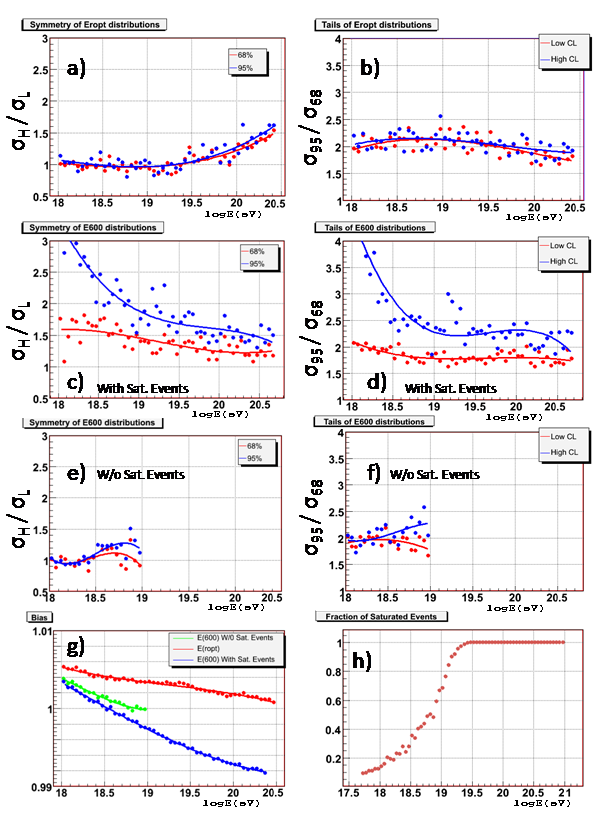}
  \end{center}
  \vspace{-0.5cm}
  \caption{From a) to f): Fist row is for E($r_{opt}$), the second row is for E(600)
  with saturated events and the third row is for E(600) without saturated events;
  left column are the symmetry of the distributions comparing their high and low
  sides (Gaussian $\sigma_H / \sigma_L =1$) and right column are the tails of the distributions
  comparing their C.L. at 68\% and 95\% (Gaussian $\sigma_{95} / \sigma_{68} =2$).
  (g) reconstructed energy bias. (h) fraction of saturated
  events.}\label{fig:rec_errors}
\end{figure}

First, using a flat spectrum from $10^{17.7}$ to $10^{21.0}$ eV we
calculated the distribution function of events contributing to each
reconstructed energy both, for $r(600)$ and $r_{opt}$. The
corresponding 68\% and 95\% confidence levels (at the low ({\bf
L\/}) and high energy sides ({\bf H\/})) for each distribution are
shown in fig. \ref{fig:rec_errors}.a-f. These distribution are very
nearly Gaussian for $r_{opt}$ (fig.\ref{fig:rec_errors}.a-b) but
skewed for $r(600)$ (fig.\ref{fig:rec_errors}.c-d). This behaviour
is somehow improved if events with saturated stations are eliminated
(fig.\ref{fig:rec_errors}.e-f), although at the high cost of
severely decreasing high energy statistics
(fig.\ref{fig:rec_errors}.h). Finally, we compare the median of the
distributions with the corresponding reconstructed energy in order
to assess the bias in each case (see, fig.\ref{fig:rec_errors}.g).
In all cases the bias is negligible. To avoid border effects, last
bins of the spectrum has been erased in the figures.

\begin{figure}
  \begin{center}
    \includegraphics[width=0.48\textwidth]{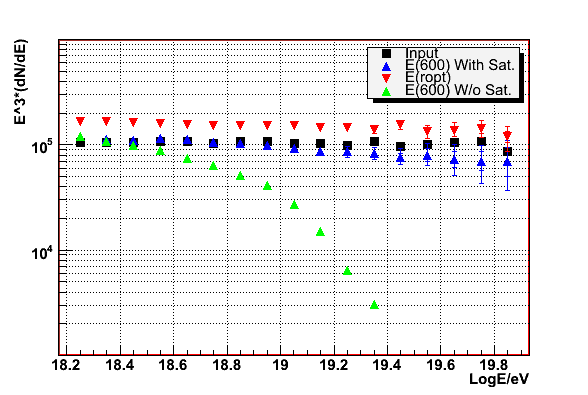}
  \end{center}
  \vspace{-0.5cm}
  \caption{Spectrum input (black) with the same statistic as AGASA and the reconstructed one
  obtained using $S_{0}(600)$ ($S(r_{opt})$) as energy estimator in blue (red). In green is the spectrum obtained
  using $S_{0}(600)$ but rejecting saturated events. The vertical axis is multiplied by $E^{3}$.
  Error bars are the CL at 68 and 95\% with 112 spectrums.}\label{fig:rec_PLspec}
\end{figure}

Figure \ref{fig:rec_PLspec} shows a reconstructed power law spectrum
from $10^{18.0}$ to $10^{20.1}$. The slope of the spectrum is better
reconstructed using $S(r_{opt})$ than $S(600)$. Again a considerable
improvement is obtained by neglecting events with saturated
stations, but at a high statistical cost. Furthermore, using
$S_{0}(600)$ around 11\% of the events are reconstructed outside of
the input bounds (most of them are events in the lower energy bins).
However, in the case of $S(r_{opt})$ this is reduced to $\sim 1$\%
of the events.

It is important to emphasize two things. First, in both cases,
$S_{\theta}(600)$ and $S(r_{opt})$, the $\chi^2/ndof$ of the
corresponding fits are very good (at the level of $10^{-2}$).
Second, in the case of $S_{\theta}(600)$, the energy reconstruction
is very bad for events with one or more saturated stations and
energy below $10^{18.5}\; eV$ so they were rejected to improve
reconstruction. This problem does not happen with $S(r_{opt})$,
highlighting a major advantage of the latter method.

\begin{figure}
  \begin{center}
    \includegraphics[width=0.48\textwidth]{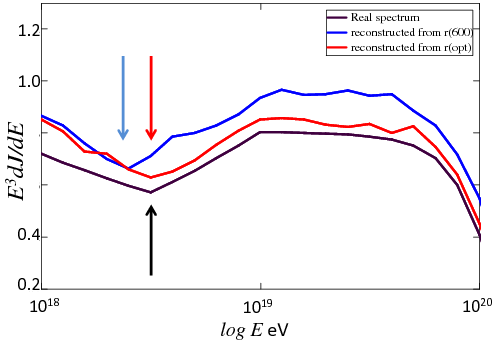}
  \end{center}
  \vspace{-0.5cm}
  \caption{Input structured spectrum (black) and the reconstructed
  spectra using $S_{0}(600)$ (blue) and $S(r_{opt})$ (red) as energy
  estimator note the position of the ankle and the GZK
  transition.}\label{fig:toy_spec}
\end{figure}

Figure \ref{fig:toy_spec} shows the effects of both energy
reconstructions when applied to a realistically structured spectrum
(black curve), that has a an ankle, a GZK modulation, observed by an
array with $\tanh$ acceptance that attains full efficiency $\lesssim
10^{18}$ eV. It can be seen that while the $S(r_{opt})$ method
fairly reproduces the impinging spectrum, the $S_{\theta}(600)$
method changes the position of the ankle and smoothes the GZK
transition.

Therefore, using an optimum distance, calculated for each individual
shower to estimate primary energy is a simple procedure to improve
the reliability of the calculated spectrum. Additionally, the
$r_{opt}$ strategy minimizes the dead-time introduced by saturated
stations or the possible biases originated by the implementation of
algorithms designed to recover saturated signals.

\section{Acknowledgements}
G. Ros thanks the Comunidad de Madrid for a F.P.I. fellowship and
the HELEN program. This work is partially supported by Spanish
grants FPA2006-12184-C02, CAM/UAH2005/071, CCG06-UAH/ESP-0397, and
Mexican PAPIIT/CIC, UNAM.


\begin{thebibliography}{99}


\bibitem{Hillas}
 A. M. Hillas.
 \emph{Acta Phys. Acad. Sci. Hung.}, 29, Suppl. 3, 355 (1970).

\bibitem{AGASASpectrum}
 Takeda et al.
 \emph{Ap. Phys.} 19 (2003) 447.

\bibitem{GustavoAlgorithm}
 G. A. Medina-Tanco et al.
 \emph{Proceedings of th 29th ICRC}, Pune, 7, (2005) 43.

\bibitem{Auger}
 Pierre Auger Collaboration. \emph{NIM}, 523, (2004) 50.

\bibitem{M.C.Medina}
 M. C. Medina et al.
 \emph{NIM A}, 566, (2006) 302.

\bibitem{GRosbienal}
 G. Ros et al.
 \emph{Res\'umenes de la XXXI Reuni\'on bienal de la RSEF}, (2007).






\end{thebibliography}
\bibliographystyle{plain}

\end{document}